\documentclass[aps,pre,twocolumn,groupedaddress,showpacs]{revtex4}
\usepackage{graphicx}
\newcommand{\br}{\mathbf{r}}

\newcommand{\sep}{ \ \ \ , \ \ \ }

\newcommand{\beq}{\begin{equation}}
\newcommand{\eeq}{\end{equation}}
\newcommand{\beqn}{\begin{eqnarray}}
\newcommand{\eeqn}{\end{eqnarray}}
\newcommand{\pp}{\partial}
\newcommand{\dd}{{\rm d}}
\newcommand{\ee}{{\rm e}}

\newcommand{\eq}{Eq.\ }

\begin{document}

\title{Density functional methods for polymers: a coil-globule transition case study
}
\author{Chiu Fan \surname{Lee}}
\affiliation{Physics Department, Clarendon Laboratory,
Oxford University, Parks Road, Oxford OX1 3PU, UK}


\date{\today}

\begin{abstract}
We consider a free energy functional on the monomer density function  that is suitable for the study of coil-globule transition. We demonstrate, with explicitly stated assumptions, why the entropic contribution is in the form of the Kullback-Leibler distance, and that the energy contribution is given by two-body and three-body terms. We then solve for the free energy analytically on a set of trial density functions, and
reproduce de Gennes' classical theory on polymer coil-globule transition.
We then discuss how our formalism can be applied to study polymer dynamics from the perspective of dynamical density function theory.
\end{abstract}
\pacs{82.35.Lr, 71.15.Mb}

\maketitle


\section{Introduction}
\begin{figure}
\caption{The distance $r$ is the measured from the monomer to the center of mass of the polymer as depicted schematically below.
}
\label{main_pic}
\begin{center}
\includegraphics[scale=.15]{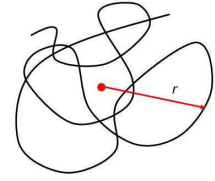}
\end{center}
\end{figure}

Polymer coil-globule transition at equilibrium has important
biological and technological importance and has thus received its well-deserved attention \cite{deGennes75,Lifshitz78,deGennes79,Kremer81, Kholodenko84,Williams81, Grosberg92a,Grosberg92b,Grosberg92c,Grosberg92d}. Here, we attempt to devise a simple formalism that treats the free energy as a functional on the monomer density. The idea of having a free energy functional is of course not new \cite{Lifshitz78,Grosberg92a}, but we believe that our contribution distinguishes itself by its simplicity, and by its emphasis on the entropic ground state. Specifically, we consider a
free energy functional of the form:
    \beq
   {\cal F}[\rho(r)] = \int \dd r f ( \rho(r))
   \eeq
   where $f$ is some function, $\rho(r)$ is the normalized monomer density function (i.e., $\int dr \rho(r) =1$) and $r$ is the Monomer-to-Center-of-Mass (MCM) distance (c.f. Fig.~\ref{main_pic})
In constructing the free energy functional, we argue that the entropic contribution should correspond to the Kullback-Leibler distance \cite{Cover91, Duda01} between $\rho(r)$ and the entropic ground state; and the energy contribution should consist of two-body term and three-body terms. We then introduce a set of trial density functions that are suitable for the coil-globule transition, and solve for the corresponding free energy analytical. We find that our approach is equivalent to de Gennes' classical theory on polymer coil-globule transition.
We then discuss how our work connects to the exciting development in dynamical density functional method \cite{Marconi99, Marconi00, Archer04}, and derive a second-order differential equation governing the dynamics of $\rho(r)$ that incorporates the entropic effect, monomer-monomer interaction, brownian motion of the monomer-monomer interaction.

\section{The free energy functional}
\label{sec:FEF}
Our aim is to construct a free energy functional that depends on the MCM distance distribution, $\rho(r)$. We firstly discuss the entropic part, and we start by writing the entropy term as:
\beq
S=-\sum_i p_i \log p_i \ ,
\eeq
where the sum is over all available states.
In a 3D regular lattice, a $N$ segments phantom chain would have a total of $\Omega=6^N$ states. Let us denote the corresponding MCM density function for each of the path by $\theta_i (r), 1 \leq i \leq \Omega$. The optimal MCM density function is then:
\beq
\rho_0 (r) = \Omega^{-1}\sum_i \theta_i(r) \ .
\eeq
We now assume that $\theta_i$ are slowly varying in $i$ and
adopt the continuous notation:
\beq
\label{rho0}
\rho_0(r) = \int_0^1 \dd s \theta(s,r)
\eeq
where $s = i/\Omega$ and is in the range $[0,1]$.
For any density function $\rho(r)$ other than $\rho_0(r)$, we have
\beq
\label{theta1}
\rho (r) =   \Omega \int_0^1 \dd s q(s) \theta(s,r) \ ,
\eeq
where $\int ds q(s) = \Omega^{-1}$, and $q(s)$ is not uniform as $\rho \neq \rho_0$ by assumption. The corresponding entropy for $\rho$ is
\beq
\Omega \int_0^1 \dd s q(s) \log(q(s)) \ .
\eeq
To construct an entropic functional $S[\rho]$ of the form:
\beq
 S[\rho(r)]=\int \dd r f(\rho(r))
 \eeq
where $f$ is some function, we need to go from the configuration enumeration index $s$ to the MCM distance $r$ (c.f. \cite{Lifshitz78}).
As far as we know, there is no known analytical expression relating $s$ and $r$. To make progress, we will now make a series of  approximations. Firstly, we assume that there exists an inverse, $\hat{\theta}$, in \eq \ref{theta1} in the sense that:
\beq
q(s) = \int \dd r \hat{\theta}(s,r) \rho(r) \ .
\eeq
We then assume that $\hat{\theta}(s,r)$ is of the form $\tilde{\theta}(s) \delta(\xi(s)- r)$ where $\xi$ is a map that takes the configuration enumeration index $s$ to the MCM distance $r$, and $\delta(.)$ is the Dirac delta function. The assumption is equivalent to saying that $q(s)$ is determined by the value of $\rho$ at the point $r=\xi(s)$ alone.
With this assumption, we have
\beq
q(s)= \tilde{\theta}(s) \rho(\xi(s)) \ ,
\eeq
which gives:
\beq
S[\rho] = \Omega\int_0^1 \dd s \tilde{\theta}(s) \rho(\xi(s)) \log[\tilde{\theta}(s) \rho  (\xi(s))] \ .
\eeq
Since we know from \eq \ref{rho0} that
\beq
1 = \tilde{\theta}(s) \rho_0(\xi(s)) \ ,
\eeq
we can solve for $\tilde{\theta}(s)$ and obtain the following:
\beq
S[\rho] = \int_0^1 \dd s \frac{\rho(\xi(s))}{\rho_0(\xi(s))} \log \left[\frac{\rho(\xi(s))}{\Omega \rho_0(\xi(s))} \right] \ .
\eeq
Let us now consider what the map $\xi$ should be. The interchange of the enumeration index and the spatial parameter can be seen as a way to weight the sum in the above integral differently as the MCM distance varies. It is intuitive to set the weight according to the magnitude of the optimal density function, $\rho_0(r)$. Namely, the difference between $\rho$ and $\rho_0$ at $r$ is given more emphasis when $\rho_0(r)$ is large. This intuition suggests that we set $ds = \rho_0(r)dr$. In other words, the map $\xi$ is defined implicitly by the following equation:
\beq
s = \int_0^{\xi(s) } \dd r' \rho_0(r')  \ .
\eeq
Now with this particular map, we have for the entropic functional the following expression:
\beq
\label{entropy}
S[\rho] =-\log \Omega +\int_0^\infty \dd r \rho(r) \log \left(\frac{\rho(r)}{ \rho_0(r)} \right) \ ,
\eeq
where we recall that $\log \Omega = N\log(6)$.
Besides the constant term, the above expression is identical to the Kullback-Leibler distance, or the relative entropy, between $\rho$ and $\rho_0$. We note that the Kullback-Leibler distance is a well studied measure in the context of information theory \cite{Cover91,Duda01} and we have also recently employed it in the context of sampling and optimization \cite{Wolpert06}.

As a functional in $\rho$, a desirable property is that one can obtain the optimal density by simply considering the zeros of the corresponding functional derivative. Now,
\beq
\frac{\delta S[\rho]}{\delta \rho}=
\log \left(\frac{\rho}{ \rho_0} \right) +1  \ .
\eeq
Setting the above expression to zero suggests that:
\beq
\rho = \ee^{-1} \rho_0 \ ,
\eeq
which is
contrary to our expectation that $\rho$ should be $\rho_0$ exactly. This is a manifestation of the fact that the functional in \eq \ref{entropy} alone is not probability preserving. Indeed, given a system with $M$ possible states, a better expression for the system's entropy is:
\beq
S=-\sum_{i=1}^{M-1} p_i \log \left( p_i\right) - \left( 1-\sum_{i=1}^{M-1}p_i \right) \log\left( 1-\sum_{i=1}^{M-1} p_i\right)\ ,
\eeq
where the unity of the sum of a probability distribution is explicitly accounted for. With large $M$  the second term in the above expression is small and as such the leading order term is $\left( 1-\sum_{i=1}^{M-1}p_i \right)$. In other words, the entropic functional can be expressed as:
\beq
\label{entropy2}
S[\rho] =-\log \Omega+1+\int_0^\infty \dd r \rho(r) \left[\log \left(\frac{\rho(r)}{ \rho_0(r)} \right) -1 \right] \ .
\eeq
With this modification, it can be checked that $\rho_0$ does correspond to the state where the functional derivative vanishes.
Note that for a probability distribution $\rho(r)$, the above modification is superfluous, but the manipulation may be necessary when considering dynamical effects. In summary,
\eq \ref{entropy2} constitutes the entropic functional that we will use here.
\\
\\
\indent
We now discuss the energy contribution in the free energy functional. We assume nearest neighbor interactions in our lattice model and we let $\eta$ be the size of the monomer. Consider a concentric spherical slice at position $r$ of thickness $\dd r$. The number of monomers in this shell
is:
\beq
 N \rho(r) \dd r\ ,
\eeq
and the number of sites available in this shell is:
\beq
 \frac{4 \pi r^2 \dd r}{\eta^3} \ .
\eeq
Therefore, the probability of any particular site being occupied is:
\beq
p(r) = \frac{N \rho(r) \eta^3}{ 4\pi r^2}\ .
\eeq
Adopting a mean-field perspective and ignoring the interaction between different shells for the time being, we have the following energy functional:
\beqn
E_1[\rho] &=& 4\epsilon \int \dd r \frac{4\pi r^2}{\eta^3} \left(\frac{N \rho(r) \eta^3}{4 \pi r^2} \right)^2
\\
&=&\epsilon \int \dd r \frac{N^2 \rho(r)^2 \eta^3}{\pi r^2} \ ,
\eeqn
where the factor 4 in front corresponds to the maximum number of neighbors allowed in our slice of volume.

The interactions between the shell at position $r$ and the shells at positions at positions $r+\eta$ and $r-\eta$ can be written as:
\beqn
&&\epsilon [p(r)p(r+\eta )+p(r-\eta )p(r)]
\\
&=&\epsilon p(r)[p(r+\eta )+p(r-\eta )-2p(r)+2p(r)]
\\
&=& \epsilon p(r) [\eta^2 p''(r) +2p(r)]
\eeqn
where $p''(r) \equiv \dd^2 p/ \dd r^2$ and $\eta$ is assumed to be small enough that the approximation by the differential operator is possible.
Writing it as an energy functional, this shell-shell interaction term becomes:
\beq
\label{E2}
E_2[\rho ] = \frac{\epsilon}{2} \int \dd r N \rho(r)
\left[ \eta^2 \frac{\pp^2}{\pp r^2}
 \left(\frac{N \rho(r) \eta^3}{4 \pi r^2} \right) +
 \frac{N \rho(r) \eta^3}{2 \pi r^2} \right] \ ,
\eeq
where the factor $1/2$ appears to take care of the double counting.
We restrict ourselves to the case where $\eta$ is small, hence we will ignore the first term in the squared brackets in \eq \ref{E2}, which is of order $\eta^5$. The overall energy contribution to the free energy functional is then simply:
\beq
\label{two_body_E}
E[\rho]=\frac{5\epsilon}{4} \int \dd r \frac{N^2 \rho(r)^2 \eta^3}{\pi r^2} \ .
\eeq

We note that we could have arrived at the above functional by assuming that $p(r)$ is slowly varying and so that the $p''(r)$ term can be ignored; but this route would not allow us to see what the magnitude of error in our approximation is. Our consideration also allows for the possibility of investigating how the situation would change with large $\eta$, in which case the energy term will be a functional dependent on both $\rho$ and $\rho''$.

As seen by inspecting \eq \ref{two_body_E}, a two-body attractive term alone will drive the polymer to the complete collapse state at any finite temperature. This can be remedied by the volume exclusion effect. To account for such an effect, we assume that the collapse state corresponds to a ball with a uniform density, i.e., $\rho(r) \propto r^2$. This suggests that the energy functional should be of the form:
\beq
E[\rho] = \int dr \left[ -\frac{A\rho^2}{r^2} +  \frac{B\rho^3}{r^4} \right]
\eeq
where $A$ and $B$ are arbitrary parameters. The second term above clearly corresponds to a three-body repulsive term as usually introduced by hand in the literature (e.g., see \cite{Pitard98}). In our case, a three-body repulsive term arises naturally by fixing the ground state of the energy functional.

In summary, by introducing two new parameters, $w_1$ and $w_2$, our energy functional becomes:
\beq
\label{new_E}
E[\rho] =\int \dd r \left[ (w_1-\epsilon) \frac{N^2 \rho^2}{r^2}
+w_2 \frac{N^3 \rho^3}{r^4} \right] \ ,
\eeq
where the constants in the attractive two-body term are absorbed into $\epsilon$. The optimizing density function for \eq \ref{new_E} now corresponds to a uniform ball with density $2(\epsilon -w_1)/3 w_2 N$.
\\
\\
\indent
Putting the entropic and energy terms together, the full free energy functional is:
\beqn
\label{FE}
\nonumber
{\cal F}[\rho] &=& -T\log \Omega+ T
 +\int \dd r  T\rho \left[\log \left(\frac{\rho}{\rho_0} \right)-1\right]
\\
&&\ +\int \dd r \left[ (w_1 -\epsilon) \frac{N^2\rho^2}{r^2}+ w_2 \frac{N^3\rho^3}{r^4}\right] \ .
\eeqn

\begin{figure}
\caption{The circles denotes the shapes of $\rho_0(r)$ corresponding to a phantom chain with unit length segments for various $N$ obtained from numerical simulations. The solid lines depict the approximation by $\phi_{\gamma(N)}(r)$. For each $N$, $\gamma$ is found by equating the mean of $\phi_\gamma$ to the mean of $\rho_0$. It is found that $\gamma=0.0063, 0.0031, 0.0016$ for $N= 500, 1000, 2000$ respectively. In other words, viewing $\gamma$ as a function of $N$, we have $\gamma(N)\times N \sim \pi$.
}
\label{approx}
\begin{center}
\includegraphics[scale=.4]{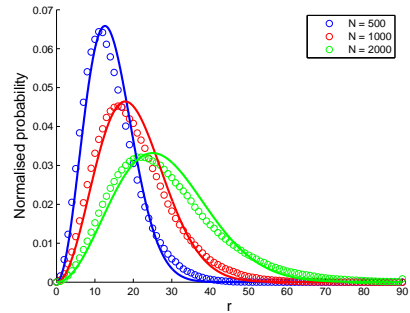}
\end{center}
\end{figure}

\section{Variational method}
We now apply the free energy functional to investigate polymer coil-globule transition. Looking at \eq \ref{FE}, we see that we firstly need to get a handle on $\rho_0(r)$. We are unaware of any analytical solution for $\rho_0(r)$. Here we adopt the simple assumption that for the a chain with $N$ segments of length $l_p$,
\beq
\rho_0(r)=\frac{4\gamma^{3/2}}{\pi^{1/2}} \left(\frac{r}{l_p}\right)^2 \ee^{-\gamma r^2/l_p^2} \ ,
\eeq
where the corresponding radius of gyration is $l_p\sqrt{3/2\gamma}$ and as such,
\beq
\label{scaling}
\sqrt{\frac{3}{2\gamma}} = \frac{R_G}{l_p} \sim N^{1/2} \ .
\eeq
Fig.~\ref{approx} illustrates that the approximation is in reasonable agreement with the density function obtained by numerical means. In particular, we find that $\gamma(N) \times N \sim \pi$.

The above consideration motivates us to introduce the set of trial density function of the form:
\beq
\label{theta}
\phi_g(r) = \frac{4g^{3/2}}{\pi^{1/2}}  \left(\frac{r}{l_p}\right)^2 \ee^{-g r^2/l_p^2}
\eeq
where $g$ is treated as the variational parameter. Since the radius of gyration corresponding to $\phi_g(r)$ is $l_p\sqrt{3/2g}$, which goes to zero as $g$ goes to infinity, this set of density functions is also suitable to describe the collapse state. With that said, we note that these trial functions do not represent well the
 minimal energy configuration because a uniformly dense ball has sharp density cutoff at the boundary while our set of trial functions have stretched tails, but we believe that this discrepancy would not be very important when the polymer is away from the completely collapse state.

Besides being capable of describing both the free and collapse states, our trial functions also have the virtue of allowing us to compute the free energy in \eq
\ref{FE} analytically. Solving for all the Gaussian integrals in the functional and writing the free energy as a function of $g$, we obtain:
\beqn
\nonumber
{\cal F}(g) &=& T\left[
\frac{3}{2}\log \left(\frac{g}{\gamma}\right) -
\frac{3 (g-\gamma)}{2g} -\log \Omega +1\right]
\\
&& \
+(w_1-\epsilon)g^{3/2}N^2\sqrt{\frac{2}{\pi}}+w_2\frac{16g^3N^3}{3^{3/2}\pi} \ .
\eeqn
Recall that the radius of gyration, $R_G$, corresponding to $\phi_g$ is $\sqrt{3/2g}$, we rewrite the free energy in terms of $R_G$ and keep only terms that depend on $R_G$:
\beq
{\cal F}(R_G) =
-3T\log (R_G) + \frac{\gamma T R_G^2}{l_p^2}+(w_1-\epsilon)\frac{ N^2}{ R_G^3}
+w_2\frac{N^3}{ R_G^6} \ .
\eeq
As a comparison, we reproduce the free energy formula as deduced originally by de Gennes below (c.f. \eq 2.42 in \cite{Sneppen05}):
\beq
\hat{{\cal F}}(R_G) = - 3T \log(R_G)+\frac{9T R_G^2}{4l_p^2N}+\frac{3(T \hat{v} - \hat{\epsilon} )}{4}  \frac{N^2}{R_G^3}
 + \frac{\hat{v}^2 T}{4 }\frac{N^3}{ R_G^6} \ ,
\eeq
where $\hat{v}$ is the volume of the monomer and $\hat{\epsilon}$ is the attractive energy. By inspection, we see that the two formula are equivalent with the following mapping:
\beqn
\gamma \mapsto \frac{9}{4N}
&\sep &
w_1 \mapsto \frac{3\hat{v}T}{4}
\\
\epsilon \mapsto \frac{3 \hat{\epsilon}}{4}
&\sep &
w_2 \mapsto \frac{\hat{v}^2T}{4} \ .
\eeqn
Indeed, these maps could have been anticipated by dimensional analysis on \eq \ref{scaling} and \eq \ref{W}. These correspondences show that our variational method on the free energy functional are equivalent to de Gennes' theory on coil-globule transition,
with except that the argument $R_G$ in our approach is an index for the density function $\phi_{2/3R_G^2}(r)$. Namely, unlike de Gennes' approach, the free energy is no longer defined by the EED.  Furthermore, our approach did not rely assuming that the radius of gyration is equal to the EED in the derivation.

\section{Conclusion and outlooks}
In this paper, we have introduced a free energy functional on the MCM density function, $\rho(r)$, and applied it to the study of polymer coil-globule transition.
In constructing the free energy functional, we argued that the Kullback-Leibler distance measure is, under clearly delineated assumptions, a suitable functional for the entropic part. We then solve for the free energy exactly on a set of trial density functions and found that our approach is equivalent to de Gennes' classical theory on polymer coil-globule transition.

In terms of outlook, we believe that our work makes a connection to the important development in dynamical density functional method \cite{Marconi99, Marconi00, Archer04} by allowing for the possibility of studying polymers dynamics with our formalism. For instance, for any polymer out of equilibrium at $t=0$, according to the dynamical density functional theory, the temporal evolution of the MCM density function may be described by
\beq
\label{DDFT}
\zeta \frac{\pp \rho(\br,t)}{\pp t} =  \vec{\nabla} \cdot \left[
\rho(\br, t) \vec{\nabla} \frac{\delta {\cal F}[\rho(\br, t)]}{\delta \rho(\br, t)} \right] \ ,
\eeq
where $\zeta$ is the friction constant.
 For our free energy functional, it leads to the following second-order differential equation:
\beq
\zeta \frac{\pp \rho}{\pp t} =
\frac{\pp}{\pp r} \left[
\rho \frac{\pp}{\pp r}   \left( T\log \frac{\rho}{\rho_0} +\frac{2(w_1-\epsilon )N^2 \rho}{r^2} + \frac{3w_2N^3 \rho^2}{r^4} \right)
 \right] \ .
\eeq
The above differential equation encapsulates the entropic effect, brownian motions of the monomers and the monomer-monomer interaction. We believe that it would serve to be a natural platform for the study of single polymer dynamics such as in the scenarios of polymer collapse \cite{deGennes85, Grosberg88,Dawson94, Timoshenko95,Buguin96, Klushin98,Pitard98,
Halperin00}
, translocation \cite{Sung96,  Muthukumar01,Chuang01,Kantor04,Dubbeldam07,Grosberg06,Ali06,Gopinathan07,Sakaue07,
Lee07} and adsorption \cite{O'Shaughnessy05}.

\begin{acknowledgements}
The author thanks the Glasstone Trust (Oxford) and Jesus College (Oxford) 
for financial support.
\end{acknowledgements}

\bibliography{DFT,new_adsorption}

\begin{thebibliography}{36}
\expandafter\ifx\csname natexlab\endcsname\relax\def\natexlab#1{#1}\fi
\expandafter\ifx\csname bibnamefont\endcsname\relax
  \def\bibnamefont#1{#1}\fi
\expandafter\ifx\csname bibfnamefont\endcsname\relax
  \def\bibfnamefont#1{#1}\fi
\expandafter\ifx\csname citenamefont\endcsname\relax
  \def\citenamefont#1{#1}\fi
\expandafter\ifx\csname url\endcsname\relax
  \def\url#1{\texttt{#1}}\fi
\expandafter\ifx\csname urlprefix\endcsname\relax\def\urlprefix{URL }\fi
\providecommand{\bibinfo}[2]{#2}
\providecommand{\eprint}[2][]{\url{#2}}

\bibitem[{\citenamefont{de~Gennes}(1975)}]{deGennes75}
\bibinfo{author}{\bibfnamefont{P.~G.} \bibnamefont{de~Gennes}},
  \bibinfo{journal}{J. Phys. Lett.} \textbf{\bibinfo{volume}{36}},
  \bibinfo{pages}{55} (\bibinfo{year}{1975}).

\bibitem[{\citenamefont{Lifshitz et~al.}(1978)}]{Lifshitz78}
\bibinfo{author}{\bibfnamefont{I.}~\bibnamefont{Lifshitz}}
  \bibnamefont{et~al.}, \bibinfo{journal}{Rev. Mod. Phys.}
  \textbf{\bibinfo{volume}{50}}, \bibinfo{pages}{683} (\bibinfo{year}{1978}).

\bibitem[{\citenamefont{de~Gennes}(1979)}]{deGennes79}
\bibinfo{author}{\bibfnamefont{P.~G.} \bibnamefont{de~Gennes}},
  \emph{\bibinfo{title}{Scaling Concepts in Polymer Physics}}
  (\bibinfo{publisher}{Cornell Univ. Press}, \bibinfo{address}{Ithaca},
  \bibinfo{year}{1979}).

\bibitem[{\citenamefont{Kremers et~al.}(1981)}]{Kremer81}
\bibinfo{author}{\bibfnamefont{K.}~\bibnamefont{Kremers}} \bibnamefont{et~al.},
  \bibinfo{journal}{J. Phys. A: Math. Gen.} \textbf{\bibinfo{volume}{15}},
  \bibinfo{pages}{2879} (\bibinfo{year}{1981}).

\bibitem[{\citenamefont{Kholodenko and Freed}(1984)}]{Kholodenko84}
\bibinfo{author}{\bibfnamefont{A.~L.} \bibnamefont{Kholodenko}}
  \bibnamefont{and} \bibinfo{author}{\bibfnamefont{K.~F.} \bibnamefont{Freed}},
  \bibinfo{journal}{J. Phys. A: Math. Gen.} \textbf{\bibinfo{volume}{17}},
  \bibinfo{pages}{2703} (\bibinfo{year}{1984}).

\bibitem[{\citenamefont{Williams et~al.}(1981)}]{Williams81}
\bibinfo{author}{\bibfnamefont{C.}~\bibnamefont{Williams}}
  \bibnamefont{et~al.}, \bibinfo{journal}{Ann. Rev. Phys. Chem.}
  \textbf{\bibinfo{volume}{32}}, \bibinfo{pages}{433} (\bibinfo{year}{1981}).

\bibitem[{\citenamefont{{Yu. Grosberg} and
  Kuznetsov}(1992{\natexlab{a}})}]{Grosberg92a}
\bibinfo{author}{\bibfnamefont{A.}~\bibnamefont{{Yu. Grosberg}}}
  \bibnamefont{and} \bibinfo{author}{\bibfnamefont{D.~V.}
  \bibnamefont{Kuznetsov}}, \bibinfo{journal}{Macromolecules}
  \textbf{\bibinfo{volume}{25}}, \bibinfo{pages}{1970}
  (\bibinfo{year}{1992}{\natexlab{a}}).

\bibitem[{\citenamefont{{Yu. Grosberg} and
  Kuznetsov}(1992{\natexlab{b}})}]{Grosberg92b}
\bibinfo{author}{\bibfnamefont{A.}~\bibnamefont{{Yu. Grosberg}}}
  \bibnamefont{and} \bibinfo{author}{\bibfnamefont{D.~V.}
  \bibnamefont{Kuznetsov}}, \bibinfo{journal}{Macromolecules}
  \textbf{\bibinfo{volume}{25}}, \bibinfo{pages}{1980}
  (\bibinfo{year}{1992}{\natexlab{b}}).

\bibitem[{\citenamefont{{Yu. Grosberg} and
  Kuznetsov}(1992{\natexlab{c}})}]{Grosberg92c}
\bibinfo{author}{\bibfnamefont{A.}~\bibnamefont{{Yu. Grosberg}}}
  \bibnamefont{and} \bibinfo{author}{\bibfnamefont{D.~V.}
  \bibnamefont{Kuznetsov}}, \bibinfo{journal}{Macromolecules}
  \textbf{\bibinfo{volume}{25}}, \bibinfo{pages}{1991}
  (\bibinfo{year}{1992}{\natexlab{c}}).

\bibitem[{\citenamefont{{Yu. Grosberg} and
  Kuznetsov}(1992{\natexlab{d}})}]{Grosberg92d}
\bibinfo{author}{\bibfnamefont{A.}~\bibnamefont{{Yu. Grosberg}}}
  \bibnamefont{and} \bibinfo{author}{\bibfnamefont{D.~V.}
  \bibnamefont{Kuznetsov}}, \bibinfo{journal}{Macromolecules}
  \textbf{\bibinfo{volume}{25}}, \bibinfo{pages}{1996}
  (\bibinfo{year}{1992}{\natexlab{d}}).

\bibitem[{\citenamefont{Cover and Thomas}(1991)}]{Cover91}
\bibinfo{author}{\bibfnamefont{T.}~\bibnamefont{Cover}} \bibnamefont{and}
  \bibinfo{author}{\bibfnamefont{J.}~\bibnamefont{Thomas}},
  \emph{\bibinfo{title}{Elements of Information Theory}}
  (\bibinfo{publisher}{Wiley-Interscience}, \bibinfo{address}{New York},
  \bibinfo{year}{1991}).

\bibitem[{\citenamefont{Duda et~al.}(2001)\citenamefont{Duda, Hart, and
  Stork}}]{Duda01}
\bibinfo{author}{\bibfnamefont{R.~O.} \bibnamefont{Duda}},
  \bibinfo{author}{\bibfnamefont{P.~E.} \bibnamefont{Hart}}, \bibnamefont{and}
  \bibinfo{author}{\bibfnamefont{D.~G.} \bibnamefont{Stork}},
  \emph{\bibinfo{title}{Pattern Classification}} (\bibinfo{publisher}{Wiley},
  \bibinfo{address}{New York}, \bibinfo{year}{2001}), \bibinfo{edition}{2nd}
  ed.

\bibitem[{\citenamefont{{Marini Bettolo Marconi} and
  Tarazona}(1999)}]{Marconi99}
\bibinfo{author}{\bibfnamefont{U.}~\bibnamefont{{Marini Bettolo Marconi}}}
  \bibnamefont{and} \bibinfo{author}{\bibfnamefont{P.}~\bibnamefont{Tarazona}},
  \bibinfo{journal}{J. Chem. Phys.} \textbf{\bibinfo{volume}{110}},
  \bibinfo{pages}{8032} (\bibinfo{year}{1999}).

\bibitem[{\citenamefont{{Marini Bettolo Marconi} and
  Tarazona}(2000)}]{Marconi00}
\bibinfo{author}{\bibfnamefont{U.}~\bibnamefont{{Marini Bettolo Marconi}}}
  \bibnamefont{and} \bibinfo{author}{\bibfnamefont{P.}~\bibnamefont{Tarazona}},
  \bibinfo{journal}{J. Phys.: Condens. Matter} \textbf{\bibinfo{volume}{12}},
  \bibinfo{pages}{A413} (\bibinfo{year}{2000}).

\bibitem[{\citenamefont{Archer and Evans}(2004)}]{Archer04}
\bibinfo{author}{\bibfnamefont{A.~J.} \bibnamefont{Archer}} \bibnamefont{and}
  \bibinfo{author}{\bibfnamefont{R.}~\bibnamefont{Evans}}, \bibinfo{journal}{J.
  Chem. Phys.} \textbf{\bibinfo{volume}{121}}, \bibinfo{pages}{4246}
  (\bibinfo{year}{2004}).

\bibitem[{\citenamefont{Wolpert and Lee}(2006)}]{Wolpert06}
\bibinfo{author}{\bibfnamefont{D.~H.} \bibnamefont{Wolpert}} \bibnamefont{and}
  \bibinfo{author}{\bibfnamefont{C.~F.} \bibnamefont{Lee}},
  \bibinfo{journal}{Europhysics Letters} \textbf{\bibinfo{volume}{76}},
  \bibinfo{pages}{353} (\bibinfo{year}{2006}).

\bibitem[{\citenamefont{Pitard and Orland}(1998)}]{Pitard98}
\bibinfo{author}{\bibfnamefont{E.}~\bibnamefont{Pitard}} \bibnamefont{and}
  \bibinfo{author}{\bibfnamefont{H.}~\bibnamefont{Orland}},
  \bibinfo{journal}{Europhysics Letters} \textbf{\bibinfo{volume}{41}},
  \bibinfo{pages}{467} (\bibinfo{year}{1998}).

\bibitem[{\citenamefont{Sneppen and Zocchi}(2005)}]{Sneppen05}
\bibinfo{author}{\bibfnamefont{K.}~\bibnamefont{Sneppen}} \bibnamefont{and}
  \bibinfo{author}{\bibfnamefont{G.}~\bibnamefont{Zocchi}},
  \emph{\bibinfo{title}{Physics in Molecular Biology}}
  (\bibinfo{publisher}{Cambridge University Press},
  \bibinfo{address}{Cambridge}, \bibinfo{year}{2005}).

\bibitem[{\citenamefont{de~Gennes}(1985)}]{deGennes85}
\bibinfo{author}{\bibfnamefont{P.~G.} \bibnamefont{de~Gennes}},
  \bibinfo{journal}{J. Phys. (Paris) Lett.} \textbf{\bibinfo{volume}{46}},
  \bibinfo{pages}{L639} (\bibinfo{year}{1985}).

\bibitem[{\citenamefont{{Yu. Grosberg} et~al.}(1988)}]{Grosberg88}
\bibinfo{author}{\bibfnamefont{A.}~\bibnamefont{{Yu. Grosberg}}}
  \bibnamefont{et~al.}, \bibinfo{journal}{J. Phys. (Paris)}
  \textbf{\bibinfo{volume}{49}}, \bibinfo{pages}{2095} (\bibinfo{year}{1988}).

\bibitem[{\citenamefont{Dawson et~al.}(1994)}]{Dawson94}
\bibinfo{author}{\bibfnamefont{K.~A.} \bibnamefont{Dawson}}
  \bibnamefont{et~al.}, \bibinfo{journal}{Nuovo Cimento D}
  \textbf{\bibinfo{volume}{16}}, \bibinfo{pages}{675} (\bibinfo{year}{1994}).

\bibitem[{\citenamefont{Timoshenko et~al.}(1995)}]{Timoshenko95}
\bibinfo{author}{\bibfnamefont{E.~G.} \bibnamefont{Timoshenko}}
  \bibnamefont{et~al.}, \bibinfo{journal}{Phys. Rev. E}
  \textbf{\bibinfo{volume}{51}}, \bibinfo{pages}{492} (\bibinfo{year}{1995}).

\bibitem[{\citenamefont{Buguin et~al.}(1996)}]{Buguin96}
\bibinfo{author}{\bibfnamefont{A.}~\bibnamefont{Buguin}} \bibnamefont{et~al.},
  \bibinfo{journal}{C. R. Acad. Sci. Paris, Ser. II b}
  \textbf{\bibinfo{volume}{322}}, \bibinfo{pages}{741} (\bibinfo{year}{1996}).

\bibitem[{\citenamefont{Klushin}(1998)}]{Klushin98}
\bibinfo{author}{\bibfnamefont{L.~I.} \bibnamefont{Klushin}},
  \bibinfo{journal}{J. Chem. Phys.} \textbf{\bibinfo{volume}{108}},
  \bibinfo{pages}{7917} (\bibinfo{year}{1998}).

\bibitem[{\citenamefont{Halperin and Goldhart}(2000)}]{Halperin00}
\bibinfo{author}{\bibfnamefont{A.}~\bibnamefont{Halperin}} \bibnamefont{and}
  \bibinfo{author}{\bibfnamefont{P.~M.} \bibnamefont{Goldhart}},
  \bibinfo{journal}{Phys. Rev. E} \textbf{\bibinfo{volume}{61}},
  \bibinfo{pages}{565} (\bibinfo{year}{2000}).

\bibitem[{\citenamefont{Sung and Park}(1996)}]{Sung96}
\bibinfo{author}{\bibfnamefont{W.}~\bibnamefont{Sung}} \bibnamefont{and}
  \bibinfo{author}{\bibfnamefont{P.~J.} \bibnamefont{Park}},
  \bibinfo{journal}{Phys. Rev. Lett.} \textbf{\bibinfo{volume}{77}},
  \bibinfo{pages}{783} (\bibinfo{year}{1996}).

\bibitem[{\citenamefont{Muthukumar}(2001)}]{Muthukumar01}
\bibinfo{author}{\bibfnamefont{M.}~\bibnamefont{Muthukumar}},
  \bibinfo{journal}{Phys. Rev. Lett.} \textbf{\bibinfo{volume}{86}},
  \bibinfo{pages}{3188} (\bibinfo{year}{2001}).

\bibitem[{\citenamefont{Chuang et~al.}(2001)}]{Chuang01}
\bibinfo{author}{\bibfnamefont{J.}~\bibnamefont{Chuang}} \bibnamefont{et~al.},
  \bibinfo{journal}{Phys. Rev. E} \textbf{\bibinfo{volume}{65}},
  \bibinfo{pages}{011802} (\bibinfo{year}{2001}).

\bibitem[{\citenamefont{Kantor and Kardar}(2004)}]{Kantor04}
\bibinfo{author}{\bibfnamefont{Y.}~\bibnamefont{Kantor}} \bibnamefont{and}
  \bibinfo{author}{\bibfnamefont{M.}~\bibnamefont{Kardar}},
  \bibinfo{journal}{Phys. Rev. E} \textbf{\bibinfo{volume}{69}},
  \bibinfo{pages}{021806} (\bibinfo{year}{2004}).

\bibitem[{\citenamefont{Dubbeldam et~al.}(2007)}]{Dubbeldam07}
\bibinfo{author}{\bibfnamefont{J.~L.~A.} \bibnamefont{Dubbeldam}}
  \bibnamefont{et~al.}, \bibinfo{journal}{Europhys. Lett.}
  \textbf{\bibinfo{volume}{79}}, \bibinfo{pages}{18002} (\bibinfo{year}{2007}).

\bibitem[{\citenamefont{{Yu. Grosberg} et~al.}(2006)}]{Grosberg06}
\bibinfo{author}{\bibfnamefont{A.}~\bibnamefont{{Yu. Grosberg}}}
  \bibnamefont{et~al.}, \bibinfo{journal}{Phys. Rev. Lett.}
  \textbf{\bibinfo{volume}{96}}, \bibinfo{pages}{228105}
  (\bibinfo{year}{2006}).

\bibitem[{\citenamefont{Ali et~al.}(2006)}]{Ali06}
\bibinfo{author}{\bibfnamefont{I.}~\bibnamefont{Ali}} \bibnamefont{et~al.},
  \bibinfo{journal}{Phys. Rev. Lett.} \textbf{\bibinfo{volume}{96}},
  \bibinfo{pages}{208102} (\bibinfo{year}{2006}).

\bibitem[{\citenamefont{Gopinathan and Kim}(2007)}]{Gopinathan07}
\bibinfo{author}{\bibfnamefont{A.}~\bibnamefont{Gopinathan}} \bibnamefont{and}
  \bibinfo{author}{\bibfnamefont{Y.~W.} \bibnamefont{Kim}},
  \bibinfo{journal}{Phys. Rev. Lett.} \textbf{\bibinfo{volume}{99}},
  \bibinfo{pages}{228106} (\bibinfo{year}{2007}).

\bibitem[{\citenamefont{Sakaue}(2007)}]{Sakaue07}
\bibinfo{author}{\bibfnamefont{T.}~\bibnamefont{Sakaue}},
  \bibinfo{journal}{Phys. Rev. E} \textbf{\bibinfo{volume}{69}},
  \bibinfo{pages}{021806} (\bibinfo{year}{2007}).

\bibitem[{\citenamefont{Lee}(2007)}]{Lee07}
\bibinfo{author}{\bibfnamefont{C.~F.} \bibnamefont{Lee}},
  \bibinfo{journal}{E-print: arXiv:0708.5764}  (\bibinfo{year}{2007}).

\bibitem[{\citenamefont{O'Shaughnessy and Vavylonis}(2005)}]{O'Shaughnessy05}
\bibinfo{author}{\bibfnamefont{B.}~\bibnamefont{O'Shaughnessy}}
  \bibnamefont{and}
  \bibinfo{author}{\bibfnamefont{D.}~\bibnamefont{Vavylonis}},
  \bibinfo{journal}{J. Phys.: Condens. Matter} \textbf{\bibinfo{volume}{17}},
  \bibinfo{pages}{R63} (\bibinfo{year}{2005}).

\end{thebibliography}
\end{document}